\documentclass[prl,aps, twocolumn]{revtex4}

\usepackage{graphicx, epsfig, fancybox}
\usepackage{amssymb,amsmath}

\def \a{\alpha}    \def \b{\beta}      
\def \e{\epsilon}       
       \def \s{\sigma}   \def \o{\omega}

\def \O{\Omega}          \def \D{\Delta}

\def \del{\partial}    
\def \hf{\tfrac{1}{2}}

\def\lba{\left(}    \def\rba{\right)}
\def\lbc{\left[}    \def\rbc{\right]}

\newcommand{\ket}[1]{\left|{#1}\right.\rangle}

\newcommand{\vp}{{\bf p}}  
\newcommand{\vq}{{\bf q}}  
\newcommand{\vk}{{\bf k}}  
\renewcommand{\vr}{{\bf r}}

\newcommand{\hc}{\hat{c}}  \newcommand{\hcd}{\hat{c}^\dag}

\begin{document}

\title{Pairing of a trapped resonantly-interacting fermion mixture
  with unequal spin populations}

\author{Masudul Haque} 

\author{H.~T.~C.~Stoof}

\affiliation{Institute for Theoretical Physics, Utrecht University,
Leuvenlaan 4, 3584 CE Utrecht, the Netherlands}

\date{\today}

%
%

\begin{abstract}

We consider the phase separation of a trapped atomic mixture of
fermions with unequal spin populations near a Feshbach resonance.  In
particular, we determine the density profile of the two spin states
and compare with the recent experiments of Partridge \emph{et al.}
\cite{Hulet_unequal-popln-expt}.  Overall we find quite good
agreement.  We identify the remaining discrepancies and pose them as
open problems.

\end{abstract}

\maketitle

\emph{Introduction} --- The usual Bardeen-Cooper-Schrieffer (BCS)
theory of Bose-Einstein condensation (BEC) of fermion pairs requires
the populations of the two species involved in the s-wave pairing to
be equal.  For a long time, therefore, theorists have discussed
fermionic pairing when the species densities are unequal, and several
proposals for the ground state have been put forward \cite{FFLO,
Sarma_1964, Sedakrian_Deformed-FS-pairing}.  Experimentally, however,
such superfluid states with unequal densities have remained elusive.

After several years of experimental studies of the BEC-BCS crossover
with equal spin population, experiments with ultracold atoms have very
recently also turned to studying superfluidity with unequal
populations \cite{Hulet_unequal-popln-expt,
Ketterle_unequal-popln-expt}.  The basic idea is to load a trap with
an unequal population of two hyperfine states of $^6$Li and tune the
bias magnetic field close to a Feshbach resonance.  It turns out that
the physics of an unequal-population Fermi mixture in a trap is rather
different from the uniform case.  the dominant characteristic of the
measured density profiles appears to be a phase separation, with an
equal-density BCS phase in an interior core, and an outer shell
consisting mostly of the majority species.  If there are additional
two-component phases, for instance Fulde-Ferrel-Larkin-Ovchinikov
(FFLO) phases, they are confined to a shell-shaped region between the
outer majority shell and the inner BCS-like core.

Partridge \emph{et al.} have reported in-situ measurements of the
density profiles of the two states \cite{Hulet_unequal-popln-expt},
while Zwierlein \emph{et al.}  report on measurements after expansion
\cite{Ketterle_unequal-popln-expt}.  The former experiments
are performed close enough to the Feshbach resonance that they may be
regarded as being in the unitarity limit, i.e., in the limit that the
interaction strength $g$ is effectively infinite so it does not
provide an energy scale to the problem.  In this Letter we concentrate
on the data from Ref.~\cite{Hulet_unequal-popln-expt} and limit
ourselves therefore to the unitarity region.  We do not consider here
the data of Ref.~\cite{Ketterle_unequal-popln-expt} on rotating
fermion gases, nor do we deal with the issues arising from expansion
after the trap is switched off.

We present a zero-temperature analysis of the phase separation using a
local density approximation (LDA) and a BCS ansatz for the many-body
wavefunction.  The Feshbach resonance is treated using a
single-channel description, because the closed-channel component of
the Cooper pair wavefunctions is small in the crossover region for the
extremely broad Feshbach resonance that is being used in the
experiment \cite{Hulet_Z-measurement, RomansStoof_Z_PRL}.
Based on an analysis of the uniform case at unitarity, we give simple
arguments for the occurrence of phase separation, and to identify the
surface that surrounds the BCS phase.  We then calculate the
majority and minority density profiles within the BCS ansatz, and
compare with experimental profiles.


\emph{BCS ansatz for the unitarity regime} ---
We first examine pairing at unitarity with unequal chemical potentials
for a homogeneous mixture.  Since we will treat the trapped case in
LDA, the results from this analysis can be used locally for any point
in the trap.
We are interested in the $g\to\infty$ limit of the Hamiltonian
\[
\hat{H} ~=~ \sum_{\vk,\s} (\e_{\vk} - \mu_{\s})\,
\hcd_{\vk,\s}\hc_{\vk,\s}
~+~ g \sum_{\vp,\vq,\vk} \hat{c}_{\vp+\vk,1}^\dag \hat{c}_{\vq-\vk,2}^\dag
\hat{c}_{\vq,2} \hat{c}_{\vp,1} 
\, .
\]
The index $\s$ runs over the two hyperfine states of $^6$Li, denoted
by $\ket{1}$ and $\ket{2}$.  The masses are the same, so
$\e_{\vk}=\hbar^2\vk^2/2m$ for both species, but the chemical potentials
are different, i.e., $ \mu_{1,2} \equiv \mu \pm h\,$, so that it is
possible to have unequal densities $n_{1,2} \equiv n\pm m\,$.

We use the BCS wavefunction as an ansatz for the paired ground state.
This corresponds to using the following decomposition for the
interaction term: $\D \sum_{\vk} \lbc \hcd_{\vk,1}\hcd_{-\vk,2} +
\hc_{-\vk,2}\hc_{\vk,1} \rbc - \D^2/g\,$.  We restrict ourselves to
zero-momentum pairing, because the experimental data do not indicate
the presence of an FFLO state \cite{data-from-Randy}, and because
two-channel calculations suggest that FFLO states are not stable close
to the resonance \cite{Radzihovsky_asymmetric-pairing_2channel}.
At unitarity, the BCS ansatz is best understood as a variational
approach as opposed to a mean-field approximation.  For the case of an
equal-density mixture, this approach has been shown to be a reasonable
method of interpolating between the BCS and BEC limits.  We employ the
same philosophy here and extract information using the BCS expressions
for the number density, energy density and the gap equation.
Fortunately, improved information about the equal-density case is
available from Monte-Carlo simulations
\cite{Georgini_crossover-MonteCarlo_PRL04, CarlsonReddy_PRL05,
Carlson-etal_PRL03} and may be used to improve our trap calculations.

For the equal-density case, and within the single-channel assumption,
the many-body system at resonance is universal in the sense that the
only energy scale in the problem is that set by the density, i.e., the
Fermi energy of the corresponding free gas $\e_{\rm F}$.  The BCS
ansatz gives the energy of the resonantly interacting system to be
${0.59}$ times $\e_{\rm F}$.  This number is called $1+\beta$, with
the universal number $\beta$ being known from Monte Carlo calculations
to be $\beta\simeq-0.58$ \cite{Georgini_crossover-MonteCarlo_PRL04,
CarlsonReddy_PRL05, Carlson-etal_PRL03}, to which the BCS value
$\beta\simeq-0.41$ should be regarded as an approximation.  In the
strong-coupling region, the pairing gap is of the order of the
chemical potential, instead of being exponentially suppressed as in
the weak-coupling regime.  Within the BCS ansatz $\D_0 \simeq 1.16\mu
\simeq 0.68\e_{\rm F}$, while Monte-Carlo calculations give $\D_0 \simeq
0.84\e_{\rm F}$ \cite{CarlsonReddy_PRL05}.

With differing chemical potentials, i.e., $h\neq{0}$, the
quasiparticle energy spectrum of the BCS ansatz has two branches
$E_{\vk,\pm} = E_\vk \pm h\,$, with $E_\vk = \sqrt{\D^2+ \xi_{\vk}^2}$
\cite{Houbiers}.  For $h>\D$, the lower branch becomes negative in the
momentum interval $(k_1,k_2)$, with $k_{1,2} = \mu \mp
\sqrt{h^2-\D^2}$.  In that case we need to fill up the negative energy
modes to construct the lowest-energy ground state.  As a result, the
thermodynamic potential becomes
\begin{multline}
\O ~=~ \frac{1}{V}\sum_{\vk} 
\lbc (\e_\vk -\mu -  E_\vk) + \frac{\D^2}{2\e_\vk} \rbc 
~-~ \frac{\D^2}{g}
\\
~+~ \Theta(h-\D)\, \sum_{k_1<|\vk|<k_2} \, E_{\vk,-}
\;\; .
\end{multline}
The $\D^2/2\e_\vk$ and $\D^2/g$ terms are required to remove the usual
ultraviolet divergence.  Note that the last term is the only place
where $h$ enters.  The gap equation is given by the condition for the
extrema of the thermodynamic potential $\O(\mu,h,\D)$ i.e.,
$\del\O/\del\D =0$.  The density average and difference may
be calculated as $n = - \del\O/\del\mu$ and $m = - \del\O/\del{h}$.
The resulting expressions also have contributions from the $(k_1,k_2)$
shell in addition to the usual BCS contributions.

The ground state of the system is the absolute minimum of $\O$.  For
equal chemical potentials, the function $\O(\mu,0,\D)$ has a maximum
at $\D=0$ and a minimum at the equal-density gap $\D_{0}$.  As $h$ is
increased, there is a certain value, $h=\D_{0}/\a_1$, at which the
$\D=0$ extremum becomes a minimum and there is an intermediate
maximum.  This maximum corresponds to an unequal-density paired
solution of the gap equation, commonly known as the Sarma phase
\cite{Sarma_1964}, which is thus unstable.  At some higher value of
the chemical potential difference, $h=\D_{0}/\a_2$, the $\D=0$ solution
becomes the \emph{global} minimum so that the normal state is more
stable than the paired state.
For weak coupling ($g^{-1}\to-\infty$) the special values of $h$ are
$h = \D/2$ and $h = \D/\sqrt2 \simeq \D/1.414$.  Moving towards the
resonance from the BCS regime, we find that the values of $\a_1$ and
$\a_2$ change only slightly within the BCS ansatz.  The change in
$\a_1$ is smaller than 1\%, while we find that $\a_2$ evolves to about
$1.44$ at unitarity, i.e., a change of about 2\%.  The value of $\a_2$
at infinite coupling is a universal number and our value $\a_2 \approx
1.44$ is the approximation to this universal number within the BCS
ansatz.

Knowing that the Sarma phase is a maximum of the thermodynamic
potential in the uniform case, it is clear that we will not have this
phase in the trap within the local-density approximation.  Thus,
barring more exotic pairing mechanisms, we only have to consider
normal phases and equal-density BCS phases.

\begin{figure}
\centering
\includegraphics[width=0.6\columnwidth]{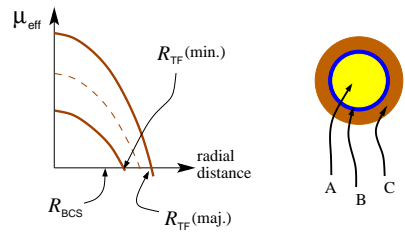}
\caption{\label{fig_cartoons} 
Left: argument for phase separation from consideration of the local
effective chemical potentials.  Shown are $\mu_{\rm eff}(r)$ (dashed)
and $\mu_{\rm eff}(r)\pm{h}$ (solid).  Right: phase separation, after
the $x$ and $y$ coordinates have been scaled to make the trap look
spherically symmetric.  Here A is the BCS core, C is the outer
majority shell, B is the intermediate shell of bi-component phase
which we treat as two non-interacting ideal gases.
}
\end{figure}

\emph{Phase separation in a trap} ---
The trapping potential used in Ref.~\cite{Hulet_unequal-popln-expt} is
asymmetric and obeys $V_{\rm trap}(\vr) = \hf{m}\o_z^2z^2 +
\hf{m}\o_{\perp}^2(x^2+y^2)\,$, with $\o_z = 2\pi\times$ (7.2 Hz) and
$\o_{\perp} = 2\pi\times$ (350 Hz).
We scale the spatial variables in the radial directions so that the
trap potential becomes spatially symmetric, with trapping frequency
$\o_z$ in each direction.
In LDA, the trap terms in the Hamiltonian are absorbed into the
chemical potential, so that we have effective space-dependent chemical
potentials:
\begin{equation}  \label{eq_effective-chempots}
\begin{split}
\mu^{\rm (1)}_{\rm eff}(r) &= \mu_{\rm 1} - V_{\rm trap}(r) =
(\mu+h) - {\hf}m\o_z^2 r^2
\\
\mu^{\rm (2)}_{\rm eff}(r) &= \mu_{\rm 2} - V_{\rm trap}(r) =
(\mu-h) - {\hf}m\o_z^2 r^2
\end{split}
\end{equation}
The average $\mu_{\rm eff} = \hf\lba\mu^{\rm (1)}_{\rm eff}+\mu^{\rm
  (2)}_{\rm eff}\rba$ decreases parabolically away from the center of
the trap while the difference equals $2h$ and stays constant.  Near
the center of the trap, $h$ is small compared to $\mu_{\rm eff}$ and
hence compared to $\D_{0}(\mu_{\rm eff})$.  Thus the densities are
forced to be equal and we have a BCS phase in the center.  Since
$\mu_{\rm eff}(r)$ decreases monotonically, there is some radius
$R_{\rm BCS}$ at which $\D_0(\mu_{\rm eff}) \simeq 1.16\mu_{\rm eff}$
is equal to ${\a_2}h$.  Outside this radius, a two-component normal
phase is more stable than a superfluid state.  For this phase, we
ignore interactions between the two components and treat it like two
ideal Fermi gases whose densities are determined by their different
chemical potentials.  At the Thomas-Fermi radius of the minority
($\ket{2}$) species, $R_{\rm TF}^{\rm (2)} = \sqrt{2(\mu-h)/m\o_z^2}$,
the minority density vanishes.  Outside $r= R_{\rm TF}^{\rm (2)}$ only
the majority remains.  This outer shell survives up to the majority
($\ket{1}$) Thomas-Fermi radius $R_{\rm TF}^{\rm (1)} =
\sqrt{2(\mu+h)/m\o_z^2}$.

\emph{Density profiles} ---
Given the chemical potentials $\mu\pm{h}$ at the trap center, we can
calculate the BCS core radius $R_{\rm BCS}$ and the Thomas-Fermi radii
$R_{\rm TF}^{\rm (1,2)}$ .  The densities are then given by
\begin{equation}  \label{eq_densities}
n_{\rm 1,2}(r) = 
\begin{cases} 
n_{\rm BCS}(\mu_{\rm eff}(r)) & r< R_{\rm BCS}
\\
n_{\rm N}(\mu_{\rm eff}(r)\pm{h}) & R_{\rm BCS} <r< R_{\rm TF}^{\rm (1,2)}
\\
0 & r>  R_{\rm TF}^{\rm (1,2)}
\end{cases}
\end{equation}
Here $n_{\rm BCS}(\mu) =
\sum_{\vk}\lbc1-(\e_\vk-\mu)/E_\vk\rbc/2V$ is the usual
equal-density BCS density for a single component, and $n_{\rm N}(\mu)
= (2m\mu)^{3/2}/6\pi^2$ is the ideal-gas density.  At unitarity,
$n_{\rm BCS}(\mu) = n_{\rm N}(\mu/[1+\b])$. 
To calculate the density corresponding to given total atom numbers, we
first find the chemical potentials which give $N_{1,2} = \int{d\vr}
n_{1,2}(r)$, and then use the expressions given above.

\begin{figure}
\centering
\includegraphics*[width=0.7\columnwidth]{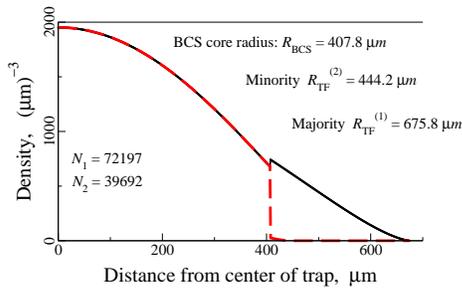}
\caption{\label{fig_unitegrated-densities} 
Majority and minority densities calculated in LDA for typical
experimental parameters of Ref.~\cite{Hulet_unequal-popln-expt}.  The
densities are expressed in units for the effective spherical trap with
trapping frequency $\o_z$ in each direction after rescaling the radial
directions.
}
\end{figure}

In Fig.~\ref{fig_unitegrated-densities}, we show density
profiles for a typical experiment of
Ref.~\cite{Hulet_unequal-popln-expt}. Experimentally, the density
itself is not accessible, but it is interesting to note some
features. 
At the BCS core radius $R_{\rm BCS}$, the majority density $n_{\rm
1}(r)$ has a discontinuity because it is determined by the BCS density
corresponding to $\mu$ on the $r<R_{\rm BCS}$ side and by the normal
density corresponding to $\mu+h$ on the $r>R_{\rm BCS}$ side.  In the
weak-coupling limit, the $n_{\rm BCS}(\mu)$ and $n_{\rm N}(\mu)$
functions are almost identical, and so the discontinuity would then
have been much more prominent.  At unitarity, however, $n_{\rm
BCS}(\mu) = n_{\rm N}(\mu/[1+\b]) = [1+\b]^{-3/2}n_{\rm N}(\mu)$, so
that $n_{\rm N}(\mu)\simeq{0.454}n_{\rm BCS}(\mu)$ for our BCS
treatment.  This significantly reduces the upward jump of the majority
density at the core edge.

Similarly, the minority density $n_{\rm 2}(r)$ has a discontinuity at
the BCS core edge as it jumps down from $n_{\rm BCS}(\mu)$ to $n_{\rm
N}(\mu-h)$.  This discontinuity is enhanced by the effect of nonzero
$\b$ in $n_{\rm N}(\mu) = [1+\b]^{3/2}n_{\rm BCS}(\mu)$.  The large
decrease of the minority density assures that the minority
Thomas-Fermi radius $R_{\rm TF}^{\rm (2)}$ is only slightly larger
than the BCS core radius $R_{\rm BCS}$, i.e., that the intermediate
two-component shell is rather thin.  In the real system, we expect
the LDA discontinuities to be smoothed out somewhat by gradient and
other corrections.  Since experimentally only spatially integrated
versions of $n_{\rm 1}(r)$ (column densities) are observed, the small
non-monotonicity of the majority density is further washed out and is
expected to be difficult to observe.

\begin{figure}
\centering
\includegraphics*[width=0.7\columnwidth]{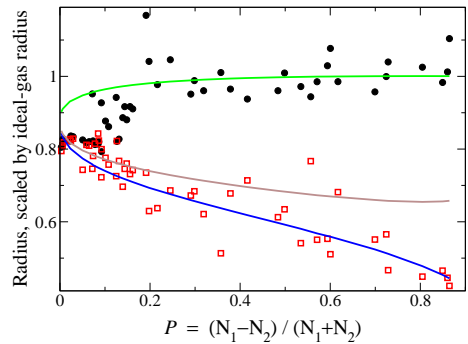}
\caption{\label{fig_R-vs-P} 
Radius versus number asymmetry.  The three solid lines from bottom to
top are calculated values of $R_{\rm BCS}$, $R_{\rm TF}^{\rm (2)}$ and
$R_{\rm TF}^{\rm (1)}$. Filled circles and empty squares are
experimental majority and minority radii from
Ref.~\cite{Hulet_unequal-popln-expt}.  The radii $R_{\rm BCS}$,
$R_{\rm TF}^{\rm (2)}$ and the experimental minority radii are scaled
by the ideal-gas Thomas-Fermi radius $R_{\rm TF}^{\rm ideal}(N_2)$ of
$N_2$ (minority) fermions.  The radius $R_{\rm TF}^{\rm (1)}$ and the
majority radii are scaled by $R_{\rm TF}^{\rm ideal}(N_1)$.
}
\end{figure}

\emph{Majority and minority radii} ---
Fig.~\ref{fig_R-vs-P} shows the evolution of the three radii in our
theory ($R_{\rm BCS}$, $R_{\rm TF}^{\rm (2)}$, $R_{\rm TF}^{\rm (1)}$)
with the number asymmetry $P=(N_1-N_2)/(N_1+N_2)$, and compares with
radii from Ref.~\cite{Hulet_unequal-popln-expt}. Measured in units of
the ideal-gas Thomas-Fermi radii corresponding to $N_{1,2}$, the
theoretical curves depend only on $P$ and not on the total number.
The experimental radii are extracted by fitting the measured column
density profiles to ideal-gas Thomas-Fermi distributions.

It is reasonable to assume that the experimental minority radii
correspond to $R_{\rm BCS}$ rather than $R_{\rm TF}^{\rm (2)}$, since
the minority occupancy in the intermediate shell is negligible, as
shown in Fig.~\ref{fig_unitegrated-densities}.  Our \emph{ab initio}
LDA calculations then explain the radius data extremely well at large
$P$.  At small $P$, the calculated radii are somewhat higher than the
experimental ones.  This is expected from the use of the BCS ansatz,
which underestimates the reduction of the paired state energy, and
hence also the reduction of the size in a trap.  We could improve our
calculation by using the Monte Carlo values $\mu \simeq 0.42\e_{\rm
F}$ and $\D_0\simeq 0.84\e_{\rm F}$.  However, to identify $R_{\rm
BCS}$ we also need the ratio $\a_2 = \D_0/h$ at which the
unequal-density normal phase becomes more stable than the
equal-density paired phase, and to the best of our knowledge $\a_2$ is
not known outside the BCS ansatz.  We have therefore opted for
consistency and used the BCS ansatz throughout.

The question remains whether there is a critical nonzero value of $P$
at which phase separation first appears.  No such feature appears in
our calculations, because with the BCS treatment of unitarity, phase
separation appears at any nonzero asymmetry, since only the
equal-density BCS phase and the normal phase are stable in this case.

\begin{figure}
\centering
 \includegraphics*[width=0.98\columnwidth]{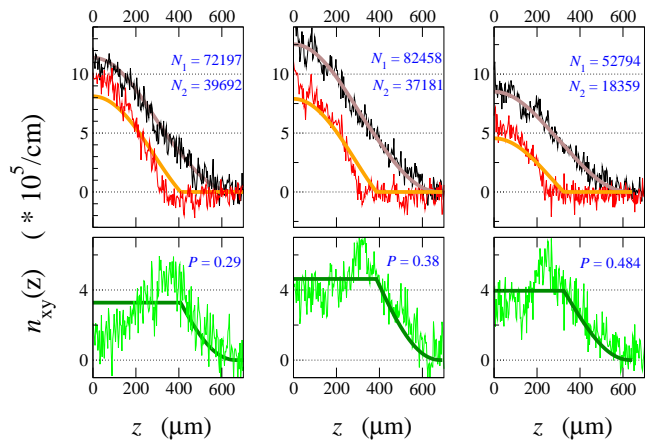} 
\caption{  \label{fig_double-integrated-densities}
Experimental axial densities as function of $z$ for several
$(N_1,N_2)$, plotted together with theoretical densities integrated
over both $x$ and $y$ directions: $\int{dx}{dy} n(x,y,z)$.  Upper
panels are the axial densities of states $\ket{1}$ and $\ket{2}$ and
the lower panels are the differences.  Note that there are no fitting
parameters.
}
\end{figure}

\emph{Axial density profiles} ---
In Fig.~\ref{fig_double-integrated-densities}, we have plotted LDA
calculations for densities with both $x$ and $y$ directions integrated
out, and compared them with column densities integrated along the $x$
direction \cite{data-from-Randy}.
 The
typical feature seems to be that the majority density profile fits
better than the minority profile, especially outside the BCS core.
This is not so surprising because, outside the core, the majority
distribution is simply the ideal-gas Thomas-Fermi distribution.
Some details of the density profiles are smoothed out because of the
double integration, but it is instructive to look at the axial density
difference.  In the integral for the axial density difference, one can
for $z<R_{\rm BCS}$ remove $z$ from both the integrand and the
integration limits, so that the theoretical axial density difference
is constant up to $z = R_{\rm BCS}$, as seen in
Fig.~\ref{fig_double-integrated-densities}.
%
%
In the experimental data, however, there is barely an extended
constant part in the difference.  Geometrically, the experimental data
indicates that the inner core is expanded radially and squeezed
axially compared to the LDA prediction.
At this point the reason for this discrepancy is not clear.  Possible
reasons could be temperature effects, nonuniversal physics beyond the
single-channel description, or the effects of nontrivial phases in the
interface region which we have not included.  Another intriguing
possibility is that, since the trap is much tighter in the radial
direction than in the axial direction, corrections to the local
density approximation might be required for the radial directions.
Chemical potentials are typically 5-15 times the energy corresponding
to radial trapping frequency, the lower end of which might be near the
limit of validity of LDA.  The discrepancy in the density difference
is an urgent issue and is presently under investigation.

\emph{Conclusion} ---
In summary, we have presented \emph{ab initio} calculations of the
density profiles for a fermion mixture near a Feshbach resonance
loaded into a trap and with unequal spin-populations. While the major
features are successfully reproduced, two major questions emerge from
our analysis.  One is the shape of the axial density difference curve
which deviates somewhat from the LDA calculation, as seen clearly in
Fig.~\ref{fig_double-integrated-densities}.  The second issue is the
possibility of having a transition from a non-phase-separated
configuration to a phase-separated configuration at a certain critical
value of the number asymmetry $P$.

We have assumed that unequal-density pairing schemes are less favored
than the two-component normal phase of the intermediate shell.  While
it is likely that the intermediate shell is too small to make a
difference in the density profiles, the issue of stability of various
unequal-density pairing schemes has not been examined thoroughly at
unitarity.  Some of these questions and issues are currently also
under investigation.

\emph{Acknowledgments} ---
We thank Randy Hulet and Wenhui Li for stimulating discussions and for
providing experimental data.  This work is supported by the Stichting
voor Fundamenteel Onderzoek der Materie (FOM) and the Nederlandse
Organisatie voor Wetenschaplijk Onderzoek (NWO).

Note: At the last stages of our work we learned of independent work
discussing issues similar to ours
\cite{Chevy_unequal-popln-thy,Duan_unequal-popln-thy}.


\begin{thebibliography}{99}


\bibitem{FFLO} P.~Fulde and R.~A.~Ferrell, Phys. Rev. {\bf 135}, A550 (1964);
\\
A.~I.~Larkin and Y.~N.~Ovchinnikov, Zh. Eksp. Teor. Fiz. {\bf 47}, 1136 (1964)
[Sov. Phys. JETP {\bf 20}, 762 (1965)]. 

\bibitem{Sarma_1964}  G.~Sarma, J. Phys. Chem. Solids {\bf 24}, 1029 (1963).

\bibitem{Sedakrian_Deformed-FS-pairing}  H.~M{\"u}ther and
  A.~Sedrakian, Phys. Rev. Lett. {\bf 88}, 252503 (2002);
Phys. Rev. C {\bf 67}, 015802 (2003).


\bibitem{Hulet_unequal-popln-expt} G.~B.~Partridge, W.~Li,
R.~I.~Kamar, Y.~A.~Liao, and R.~G.~Hulet,  cond-mat/0511752. 
%
Published online in Science,  Dec. 22, 2005.


\bibitem{Ketterle_unequal-popln-expt} M.~W.~Zwierlein, A.~Schirotzek,
  C.~H.~Schunck, and W.~Ketterle, cond-mat/0511197.  
%
Published online in Science, Dec. 22, 2005. 


\bibitem{Hulet_Z-measurement} G.~B.~Partridge, K.~E.~Strecker,
  R.~I.~Kamar, M.~W.~Jack, and R.~G.~Hulet, Phys. Rev. Lett. {\bf 95},
  020404 (2005).

\bibitem{RomansStoof_Z_PRL} M. W. J. Romans and H. T. C. Stoof ,
  Phys. Rev. Lett. {\bf 95}, 260407 (2005).  


\bibitem{data-from-Randy}  W.~Li and R.~G.~Hulet, personal
  communication.

\bibitem{Radzihovsky_asymmetric-pairing_2channel} D.~E.~Sheehy and L.~Radzihovsky
  cond-mat/0508430.  


\bibitem{Carlson-etal_PRL03}  J.~Carlson, S.~Y.~Chang,
  V.~R.~Pandharipande, and K.~E.~Schmidt, Phys. Rev. Lett. {\bf 91},
  050401 (2003).

\bibitem{Georgini_crossover-MonteCarlo_PRL04}  G.~E.~Astrakharchik,
  J.~Boronat, J.~Casulleras, and S.~Giorgini, Phys. Rev. Lett. {\bf 93}, 200404 (2004).

\bibitem{CarlsonReddy_PRL05} J.~Carlson and S.~Reddy,
  Phys. Rev. Lett. {\bf 95}, 060401 (2005).


\bibitem{Houbiers} M.~Houbiers, R.~Ferwerda, H.~T.~C.~Stoof,
  W.~I.~McAlexander, C.~A.~Sackett, and R.~G.~Hulet, Phys. Rev. A {\bf
  56}, 4864 (1997).




\bibitem{Duan_unequal-popln-thy} W.~Yi and L.-M.~Duan, cond-mat/0601006.  

\bibitem{Chevy_unequal-popln-thy} F.~Chevy, cond-mat/0601122. 




\end{thebibliography}
\end{document}